\documentclass[12pt]{article}
\usepackage{xcolor}
\usepackage{amsfonts}
\usepackage{amssymb}
\usepackage{amsmath}
\usepackage{amsthm}
\usepackage{graphicx}
\usepackage{natbib}
\usepackage{rotating}
\usepackage{setspace}
\usepackage{arydshln}
\usepackage{algorithm}
\usepackage{algpseudocode}
\usepackage{stackengine}
\usepackage{epstopdf}
\usepackage{verbatim}
\usepackage{float}
\usepackage[utf8]{inputenc}
\usepackage{booktabs}
\usepackage[font={small}]{caption}
\usepackage{pdflscape}
\usepackage{cmap}
\usepackage{varwidth}
\usepackage[flushleft]{threeparttable}
\usepackage{multirow}
\usepackage{multicol}
\usepackage{float}
\usepackage[flushleft]{threeparttable}
\usepackage{tikz}
\usetikzlibrary{arrows,positioning}
\usepackage{lipsum}

\usepackage{caption}
\usepackage{subcaption}
\usepackage{float}

\usepackage{xcolor,colortbl}
\definecolor{Cinza}{gray}{0.85}
\definecolor{Azul}{RGB}{102,175,255}
\definecolor{section.color}{HTML}{292B2E}
\definecolor{subsection.color}{HTML}{31363A}
\definecolor{subsubsection.color}{HTML}{AF5F30}
\definecolor{cor1}{HTML}{F37878}
\definecolor{cor2}{HTML}{91BDFC}

\usepackage{hyperref}
\usepackage{cleveref}
\usepackage{geometry}
\begin{document}

\begin{titlepage}
\title{The Proper Use of Google Trends in Forecasting Models}

\author{
    Marcelo C. Medeiros\footnote{Updated versions of this paper will be available on the corresponding author's webpage.}\\
	Pontificial Catholic University of Rio de Janeiro
	\and
	Henrique F. Pires\\
	Pontificial Catholic University of Rio de Janeiro
	}

\date{\small{First version: Fev 2021 \\ This version: Mar 2021}}

\maketitle
\thispagestyle{empty}

\onehalfspacing

\begin{abstract}

It is widely known that \texttt{Google Trends} has become one of the most popular free tool used by forecasters both in academics and in the private and public sectors. There are many papers, from several different fields, concluding that \texttt{Google Trends} improve forecasts' accuracy. However, what seems to be widely unknown, is that each sample of Google search data is different from the other, even if you set the same search term, data and location. This means that it is possible to find arbitrary conclusions merely by chance. This paper aims to show why and when it can become a problem and how to overcome this obstacle.

\bigskip

\noindent
\textbf{Keywords:} \texttt{Google trends}, forecasting, nowcasting, big data.
\vspace{5mm}
\end{abstract}
\end{titlepage}
\clearpage

\section{Introduction} \label{sec:introduction}

\texttt{Google Trends} is an incredible Google tool, available at no cost, that shows the most popular terms searched in the recent past. It provides access to a sample of actual search requests made to Google, allowing one to look into the interest in a particular topic from around the globe or down to city-level geography. Several studies from different used this resource mainly to improve forecast accuracy. See, for example, \citet{Askitas2009}, \cite{artola2012}, \citet{Scott2017}, \citet{Nacca2018}, and \citet{ferrara2019}.

The tool provides the frequency in which a particular term is searched for in several languages from various regions of the world. However, Google normalizes the search data to make comparisons between terms easier. This means that search results are normalized to the time and location of a query by the following process. First, each data point is divided by the total searches of the location and time range chosen by the user. Then, the resulting numbers are scaled from 0 to 100 based on the topic’s proportion to all searches on all topics.

 By doing so, \texttt{Google Trends} data represent the relative popularity. This feature avoid the situation where places with the most search volume would always be ranked the highest. Naturally, this normalization also implies that, if different regions show the same search interest for a term, it does not always mean that they actually have the same total search volumes.

All these Google Trends features described above are widely known by its users. However, what most do not seem to know (or to ignore) is the fact that almost every time one looks for a \texttt{Google Trend}'s term and downloads it, she gets a different series of numbers. The reason for this is that Google handle billions of searches per day, which means that providing access to the entire data set would be too large to process quickly. To circumvent this obstacle, only a small sample of Google searches are actually used in Google Trends. By sampling data, one can look at a dataset - generally - representative of all Google searches and that can be processed within minutes of an event happening in the real world.

However, there are some cases when this feature can actually become a real problem if not treated. We are going to show in this paper in which cases it is important to look at \texttt{Google Trend} data with some care. Moreover, we are going to show with both simulated and real data that it is possible to keep using \texttt{Google Trends} as covariates (and indeed important ones) without having the problem of your dataset not representing the actual population of the term's search for some specific period of time and location.

The article is divided as follows. Section \ref{sec:how_representative} explains the problem, highlighting how this sampling feature of \texttt{Google Trends} may be specially harmful for exercises involving nowcasts and the use of data vintages. Section \ref{sec:simulation} shows with a simulation exercise how can one strongly improves model's accuracy when using \texttt{Google Trends} as covariates. Section \ref{sec:real_data}, the last one before the conclusion, will illustrate with real data how our proposed strategy to deal with \texttt{Google Trends} can improve forecast accuracy.

\section{How representative of the actual data is each sample?}
\label{sec:how_representative}

As previously explained, Google makes available only a small sample of its search database. What most researchers and practitioners seem not to know (or to ignore) is the fact that this small sample is not always the same. In fact, it is constantly changing. This means that someone who downloads \texttt{Google Trend} data today will not download the same data tomorrow, even if she filters the same topics, languages and location.

We show with two examples of search topics (``Refined Petroleum'' and ``GDP Growth'') in two different regions (US and Brazil) that this difference in the data will be higher the less often the term is searched. Even though we do not have access to the actual number of searches for each term in each region, it is reasonable to assume that ``Refined Petroleum'' is less searched than ``GDP Growth'' and that less people use Google in Brazil than in the US (smaller population and percentage of people with access to the internet). This translates as the ``Refined Petroleum'' \texttt{Google Trends}' search data in Brazil being the more volatile and the ``GDP Growth'' data in the US the less.

It is important to note that this feature is not something that Google tries to hide. One can find information about the sampling of \texttt{Google Trends} as easy as in its \href{https://support.google.com/trends/answer/4365533?hl=en&ref_topic=6248052}{FAQ}. Nevertheless, most studies using \texttt{Google Trends} do not seem to take it into consideration. For example,  \citet{EID2009}, \citet{Vicente2015}, \citet{ferrara2019}, and \citet{BIS2020} do not even mention this fact that each \texttt{Google Trends} sampling may result in different outcomes. Others like \citet{heikk2019}, do recognize the importance of what this feature can imply, but do not seem to do anything about it in their estimations.

On the other hand, it is important to stress that there are extremely well executed papers that not only recognize the issue but also propose solutions to it. Good examples of such papers are \citet{Amuri20017}, \citet{Narita2018}, \citet{OECD2020}, \citet{Borup2020}, and \citet{Borup2021}, among others.

\begin{figure}[H]
\centering
\includegraphics[width=0.8\linewidth]{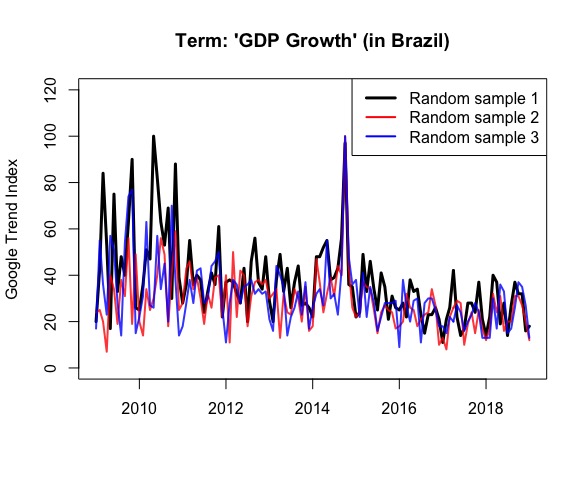}
\vspace{-1.2cm}
\caption{Three different samples of same topic and date in Brazil.}
\label{fig:gdp_growth_br}
\end{figure}

Nevertheless, the important question is how bad can these different samples be to one's forecast model? To answer it, we gathered many different samples from Google searches in the US and in Brazil between January, 2009 to January, 2019 (this way we avoid the period from both the 2008 crisis and the COVID-19 pandemic). In Figure \ref{fig:gdp_growth_br} we report the values from three different samples for the same term (``GDP Growth'') in Brazil and same timestamp. It illustrates well the fact that each \texttt{Google Trend} sample is different from the other, even when we set fixed the same topic, timestamp and region. As one can observe above, the black curve (random sample 1) does not even reach its maximum value in the same date as the other two. In case it is not very clear in the plot the lack of strong correlation between each series.

\begin{figure}[H]
\centering
\includegraphics[width=0.8\linewidth]{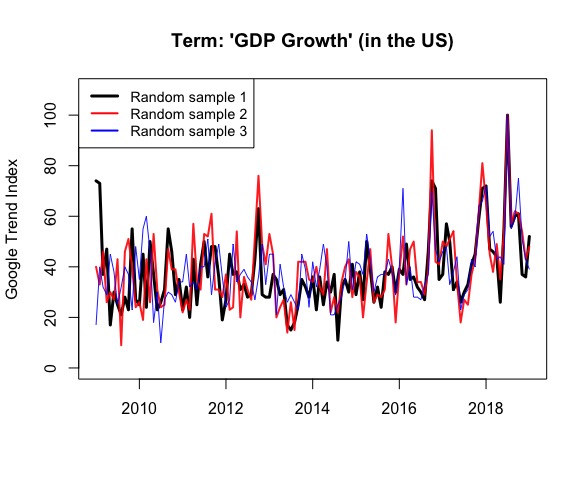}
\vspace{-1.2cm}
\caption{Three different samples of same topic and date in the US.}
\label{fig:gdp_growth_US}
\end{figure}

The correlation among series are reported in Panel (a) in Table \ref{tab:corr_gdp}. The table shows that the correlation between two different \texttt{Google Trends} sample can be as low as 0.496, even when we consider a relatively popular topic in Economics. Moreover, as the following plot and correlation matrix show, this finding is not an exclusivity of Brazil. See, Figure \ref{fig:gdp_growth_US} and Panel (b) in Table \ref{tab:corr_gdp}, where we report the results for the case of US.

\newpage

\begin{table}[!htbp]
\centering
\caption{Correlation between three different samples (S).}
\begin{minipage}{0.87\linewidth}
\begin{footnotesize}
The table reports the correlation matrix among three different samples for ``GDP Growth'' searches. Panel (a) reports the cased of Brazil and Panel (b) shows the US case.
\end{footnotesize}
\end{minipage}

\begin{tabular}{cccc}
\hline
\multicolumn{4}{c}{\underline{\textbf{Panel (a): Brazil}}}\\
 & GDP Growth - S1 & GDP Growth - S2 & GDP Growth - S3 \\
\hline \\[-1.8ex]
GDP Growth - S1 & $1$ & $0.496$ & $0.545$ \\
GDP Growth - S2 & $0.496$ & $1$ & $0.564$ \\
GDP Growth - S3 & $0.545$ & $0.564$ & $1$ \\
\\
\multicolumn{4}{c}{\underline{\textbf{Panel (b): US}}}\\
& GDP Growth - S1 & GDP Growth - S2 & GDP Growth - S3 \\
\hline \\[-1.8ex]
GDP Growth - S1 & $1$ & $0.655$ & $0.516$ \\
GDP Growth - S2 & $0.655$ & $1$ & $0.575$ \\
GDP Growth - S3 & $0.516$ & $0.575$ & $1$ \\
\hline \\[-1.8ex]
\end{tabular}
\label{tab:corr_gdp}
\end{table}

But how can a researcher use such a volatile dataset in her forecast? We propose a very simple solution in the next subsection.

\subsection{A simple way to circumvent the problem}
\label{subsec:how_cirvumvent}

The plots previously displayed\footnote{In the Appendix we show another setup in which the sampling problem with \texttt{Google Trends} may arise and should be treated with care.} illustrate well the motivation of this article, by showing that each \texttt{Google Trend} sample may be very different from each other. But does it imply that Google Trends should not be used as research tools? Not at all.

First of all, it is important to mention that very popular terms (e.g. COVID-19 in 2020) do not vary so much among different samples. However, in many situations our terms of interest are not these very popular ones. So how to overcome the possible problem shown in the plots above? The answer is very simple. By gathering many different samples and averaging across every term, one can get a more reliable time series of that term.

\begin{figure}[H]
\centering
\includegraphics[width=0.8\linewidth]{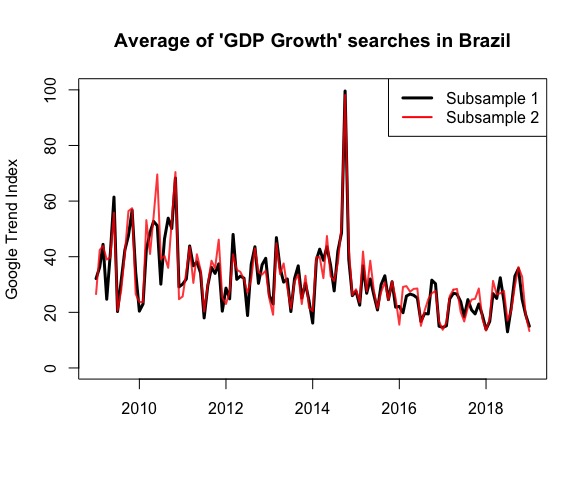}
\vspace{-1.2cm}
\caption{Averaged series using different samples in Brazil.}
\label{fig:av_gdp_growth_br}
\end{figure}

\begin{figure}[H]
\centering
\includegraphics[width=0.8\linewidth]{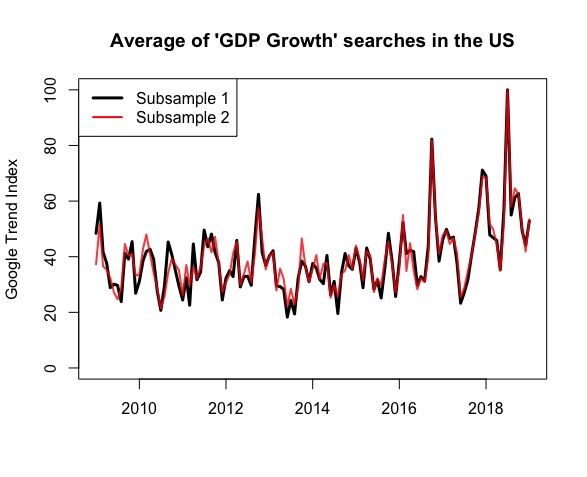}
\vspace{-1.2cm}
\caption{Averaged series using different samples in the US.}
\label{fig:av_gdp_growth_us}
\end{figure}

To illustrate how taking averages of many samples improve the series consistency, Figures \ref{fig:av_gdp_growth_br} and \ref{fig:av_gdp_growth_us} below plot the curve for the same term and same timestamp, but now comparing averages from different samples. Each curve represent the average of ``GDP Growth'' searches in seven different random samples. It is important to note that if some sample is in one average, it necessarily is not in the other one. As one can observe in the figures, taking averages strongly raise the correlation between the (averaged) samples. Besides that, as expected, the correlation between each series is a little higher in the US (0.95) than in Brazil (0.92).

\section{Simulation: Model selection improvement}
\label{sec:simulation}

The current section will exemplify through a simulation exercise how can one improve model selection performance simply by using as covariates averages of multiple \texttt{Google Trends} samples instead of a single sample. The idea here is to observe the capacity of LASSO (Least Absolute Shrinkage and Selection Operator, \citet{LASSO}) to select the correct variables (i.e., variables that indeed make part of the Data Generation Process or DGP) in two different linear regression setups.

For this exercise, we gathered 28 different \texttt{Google Trends} samples (14 for Brazil and 14 for the US) collected in different days but using the same timestamps and terms (in its respective language). We consider 20 different search terms related to Economics. When the location is Brazil, we search for terms in Portuguese, while when the location is the US, we search for the same terms in English. The searches comprehend the period from January 2009 to December 2018 (120 months). The searches are conducted at the monthly frequency.
In the first setup\footnote{In Appendix B we write the methodology in an algorithm format.}, we construct different dependent variables for each replication using only a few variables extracted from a single random sample (from the 14 downloaded) of our \texttt{Google Trends} data. Then, we run 13 different LASSO regressions, each of which having as covariates only one sample of our \texttt{Google trends} data (except the one that was used to construct the dependent variables in that iteration) to try to identify which variables (among the 20 terms in the database) are indeed in that DGP.

In the second setup we do a similar replication exercise, but in this case in each iteration we construct the dependent variables using a few variables from the average of seven (randomly selected) \texttt{Google Trends} samples to then run a single LASSO regression (using the remaining seven samples among the 14) in order to try to select the correct variables in each DGP. Finally, we compare the average (across replications) of correctly selected variables between setups 1 and 2 and conclude that using the average strongly increases model selection performance.

More specifically, in the first setup, we simulate 1,000 replications with 120 observations each. In each iteration, a random number ($s$) ranging from 1 to 14 is drawn. This will indicate which one of the 14 \texttt{Google Trends} sample we will use to generate the data. Then, we randomly pick five of the 20 search terms to form the set of relevant variables of replication $i$ that will actually be used to construct using sample $s$ ($\boldsymbol{X}_{0,i,t,s}$). Finally, for each location (Brazil or US) we generate three different dependent variables given as
\begin{equation}
\begin{split}
Y^{BR}_{i,k,t,s} &= \boldsymbol{\beta}_{i,k}^{BR}\boldsymbol{X}^{BR}_{0,i,t,s} + \epsilon^{BR}_{i,k,t,s}\\
Y^{US}_{i,k,t,s} &= \boldsymbol{\beta}_{i,k}^{US}\boldsymbol{X}^{US}_{0,i,t,s} + \epsilon^{US}_{i,k,t,s},
\end{split}
\end{equation}
where $i = 1,2, \ldots, 1000$, $k = 1,2$ or $3$ indexes the DGP and $t = 1,2, \ldots, 120$. The linear coefficients are determined as follows. $\boldsymbol\beta_{i,1}$ is a vector of integers sampled from a discrete uniform distribution in the interval $[-10,10]$; $\boldsymbol{\beta}_{i,2}$ has elements set to 1 or 2 with equal probabilities; and $\boldsymbol{\beta}_{i,3}$ is sampled from a continuous uniform distribution in the interval $[0,1]$. Finally, $\epsilon^{BR}_{i,k,t,s}$ and $\epsilon^{US}_{i,k,t,s}$ are independent and normally distributed with zero mean and are set to have the same variance as $\boldsymbol{\beta}_{i,k}^{BR}\boldsymbol{X}^{BR}_{0,i,t,s}$ and $\boldsymbol{\beta}_{i,k}^{US}\boldsymbol{X}^{US}_{0,i,t,s}$.

Then, for each replication in setup 1, after all six dependent variables are constructed, we run 13 different LASSO regressions (each one using one of the 13 remaining samples that are not used in the GDP of that iteration) to then save the topics each one of them selected as predictors. It is important to remember that each of these 13 models predicts each $Y_{i,k,t,s}$ using the full database (which has 20 variables, each one being a time series related to a different \texttt{Google Trends} term), whereas each $Y_{i,k,t,s}$ is constructed using only five random terms among the 20 possible.


As previously explained, the second setup brings a similar idea. As in setup 1, in every replication the dependent variables will be constructed using only five random terms (among the 20 topics in each database). However, instead of randomly picking only one among the 14 \texttt{Google Trends} data, in each iteration we will randomly draw seven databases, take their average for each topic and then use the five previously selected topics to construct the dependent variables. Then, after all six dependent variables are constructed, the algorithm will now run only one LASSO regression (instead of 13) which will use the average of the remaining 7 samples (the ones that are used to construct each $Y$) as covariates. Finally, the LASSO regression selects which variables (from the 20 topics available - each one being the average of that topic across the seven remaining samples) are in that iteration's DGP. The rest of the process is analogous to the first setup.

Our conclusions - which are summarized in Table \ref{tab:simulation} - are that one can highly increase model selection performance (up to 27\% in our simulations) with LASSO when using the average of multiple terms instead of a single sample from \texttt{Google Trend}'s website.

\begin{table}[!htbp] \centering
\caption{Simulation Results.}
\begin{minipage}{0.58\linewidth}
\begin{footnotesize}
The table reports the mean percentage of correctly selected variables in each model and each simulation setup.
\end{footnotesize}
\end{minipage}
\label{tab:simulation}
\begin{tabular}{@{\extracolsep{5pt}}lccccccc}
\\[-1.8ex]\hline
\hline \\[-1.8ex]
Setup & \multicolumn{1}{c}{$US^1$} & \multicolumn{1}{c}{$US^2$} & \multicolumn{1}{c}{$US^3$} & \multicolumn{1}{c}{$BR^1$} & \multicolumn{1}{c}{$BR^2$} & \multicolumn{1}{c}{$BR^3$}  \\
\hline \\[-1.8ex]
1 & 51.1 & 62.7 & 58.3 & 56.2 & 68.1 & 61.1 \\
2 & 64.72 & 73.84 & 66.82 & 67.62 & 76.08 & 68.76 \\
\hline \\[-1.8ex]
\end{tabular}
\end{table}

\section{Empirical Application}
\label{sec:real_data}

This section will advocate both for the use of \texttt{Google Trends} as a powerful prediction tool as well for the use of the average of many samples of the same search pattern as your independent variable, instead of only one sample.

During the pandemic, many authors tried to predict and understand the pattern of COVID-19 cases and death counts with all sort of models, including both epidemiological and statistical ones; see, for example,  \citet{2020arXiv200407977M}, \citet{PNAS2020}, \citet{pone2020}, \citet{2020arXiv200913484C}, and \citet{covidforecast2020}. One of the most important indicators of the actual severity of the pandemic, is the number of cases organized by the date of first symptoms felt by each person as well as the daily new deaths sorted by the day of the event (and not the date of registry). These numbers give a clear indication how the disease is evolving in real-time.

However, in most countries and specially in Brazil, the actual numbers are available with major delays (more or less 2 weeks). For some locations in Brazil, the delay can exceed a month. Part of the delay is due to the coronavirus cycle but most of it is related to the bureaucracy in the register system.\footnote{For further reference with respect to the Brazilian case, please visit the following \href{http://www.cepesp.io/rodando-os-dados-o-atraso-na-notificacao-de-casos-de-covid-19/}{link}.} This delay means that by using \texttt{Google Trends} (which is available almost in real-time) one could have a good insight of the direction in which the numbers would be moving towards two weeks in advance. We call this \emph{nowcasting}.

However, would it be possible to correctly nowcast the trend of SARS (syndrome respiratoire aigu sévère, strongly associated with COVID-19 in 2020) cases in Brazil using merely a handful of \texttt{Google Trends} topics as predictors?

\begin{figure}[H]
\centering
\includegraphics[width=0.75\linewidth]{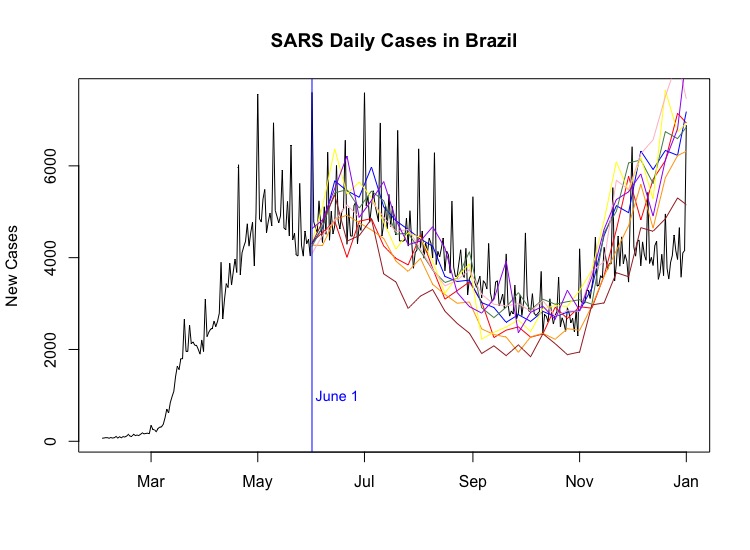}
\vspace{-1cm}
\caption{Nowcast of SARS daily cases in Brazil using only Google Trends.}
\label{fig:covid_nowcast_cases}
\end{figure}

To answer this question, we consider a simple LASSO regression to predict the trend of SARS new cases and new deaths in 2020 in Brazil. For each dependent variable (new cases and new deaths) we estimate eight LASSO regressions (each one using a different \texttt{Google Trends}' sample of the same topics, timestamps and location, but collected in different days) from February to May 2020 using only \texttt{Google Trends} data with terms related to the pandemic, such as ``COVID symptoms'', ``COVID ICU'' and ``coronavirus hospital'' as predictors\footnote{Naturally each of these terms translated to Portuguese.}. Figure \ref{fig:covid_nowcast_cases} displays the results of the nowcasting model for SRAG's new cases. The black curve in the figure represents the actual number of new SARS cases aggregated by the reported day of the first symptoms felt by each person. The colored curves represent the point forecasts of the eight different samples of the same \texttt{Google Trends} terms starting on June 1, 2020.

As it is clear in Figure \ref{fig:covid_nowcast_cases}, it would have been possible to nowcast in real-time the trend of SARS new cases using either one of the samples. In other words, by using only \texttt{Google Trends} data, one could predict that the number of new cases would continue on a high plateau until August, then it would start falling until Mid-November to then start rising again until the last day of the year. By looking at Figure \ref{fig:covid_nowcast_deaths} below, we could conclude that it would also have been possible to predict the trend of SARS new deaths by using only \texttt{Google Trends} data.

\begin{figure}[H]
\centering
\includegraphics[width=0.75\linewidth]{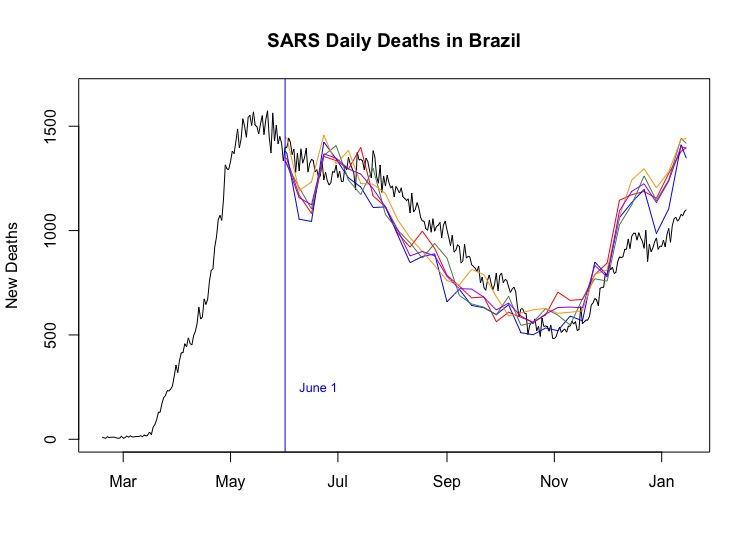}
\vspace{-1cm}
\caption{Nowcast of SARS daily deaths in Brazil using only Google Trends.}
\label{fig:covid_nowcast_deaths}
\end{figure}

But now putting on the ``black hat'', there is another thing to consider. Even though the eight predictions (using different samples) are similar, they are not the same. This is specially meaningful if we take into consideration that all of the topics used in the models described above could be considered very popular terms in the year of 2020.

To understand how using the average of the eight samples across each term would improve the forecasts results for both dependent variables, we compared the Root Mean Squared Error (RMSE) of each model (New cases and New deaths) using averaged series (Proposed Model) with the other eight models that used \texttt{only one Google Trends} sample. The results are displayed in Table \ref{tab:covid_improvement}.

\begin{table}[!htbp] \centering
\caption{Nowcasting Results.}
\begin{minipage}{0.8\linewidth}
\begin{footnotesize}
The table reports the Root Mean Squared Error (RMSE) of the nowcasting LASSO regression using the average of the \texttt{Google Trend} samples as regressors as well as the RMSE for the sample-specific models with the worst, best and average performance.
\end{footnotesize}
\end{minipage}
\begin{tabular}{lccccccc}
\hline
\hline \\
 &
 \multicolumn{1}{c}{Proposed Model} & \multicolumn{1}{c}{Worst} & \multicolumn{1}{c}{Best} & \multicolumn{1}{c}{Average} \\
\hline \\
New cases' RMSE &  888.6 & 1340.97 &  900.6 & 1058.13  \\
New deaths' RMSE &  149.2 & 175.3 &  144.4 & 159.3  \\

\hline \\[-1.8ex]
\end{tabular}
 \label{tab:covid_improvement}
\end{table}

As one can infer above, for the eight new cases' models, not only none of them had a smaller RMSE than the average, but we found out that the RMSE was up to 51\% higher in the models using only one sample. For the new death's models, the proposed method had a RMSE more than 6\% smaller than the average RMSE of the other models and more than 17\% smaller than the worse one.

These numbers above illustrate the fact that merely by chance an uninformed researcher could find herself in one of the following situations when using only a single sample of \texttt{Google Trends} in her studies. First, it could be possible that a given term is actually a good predictor of a variable of interest, but she may unfortunately find herself working with a specific random sample of \texttt{Google Trends} data that does not represent very well the true population. In this case, she may end up concluding that there is not a significant relationship between the searches on the term and the dependent variable. The other situation is the one that she may find an apparent significant relation when there is not. By averaging multiple samples, one can diminish this risk.

\section{Conclusions} \label{sec:conclusions}

This article has reinforced the usefulness of internet search data as a powerful forecast tool. However, it has also highlighted some limitations of these data, such as the lack of information on the actual volume of searches, but mainly the fact that Google Trends' index is based on a subsample that is changing all the time. We explained that this feature might become a real problem for the forecaster when dealing with less frequent searched topics and/or when filtering the query to many years in the past.

We then illustrated with two examples using both simulated and real data that simply by taking averages of many different samples with the same specifications (i.e. same topic, timestamp and location) it is possible to improve both model selection and its forecast accuracy.

\bibliographystyle{abbrvnat}

\bibliography{ref}

\newpage

\appendix

\section{Forecast models in Real Time }\label{sub:appendixA}

In 2015, \texttt{Google Trends} data was made available in real time. It is no doubt that it is helping people around the world explore the global reaction to major events. But not only that. The way \texttt{Google Trends} is constructed allows one to verify how was the search patterns in the past for a specific term and location. This feature makes possible the construction of vintages of the same search patterns, reproducing for any given day in that range, information that was available to a real-time forecaster.

That being said, it becomes clear that due to the fact that the use of real-time forecast and nowcast has been increasingly rising (some examples are \citet{rt2017}, \citet{CAMA2018}, \citet{OECD2020},  and \citet{Alex2020}), the possibility of constructing vintages of \texttt{Google Trends} is a very powerful feature of this tool. However, the same possible issue shown in previous sections may arise in a setup of real-time forecast and/or nowcast.

To illustrate the issue, we display below a series of figures. Each one contains three search pattern for the same term one month ahead. For example, the first figure below brings the time series of ``Refined Petroleum'' searches in the US from: 1) Jan, 2004 to Jan, 2014; 2) Feb, 2004 - Feb, 2014; 3) Mar, 2004 - Mar, 2014. This would be what a real-time forecaster could gather when constructing the vintages.

\begin{figure}[H]
    \centering
    \includegraphics[scale=0.5]{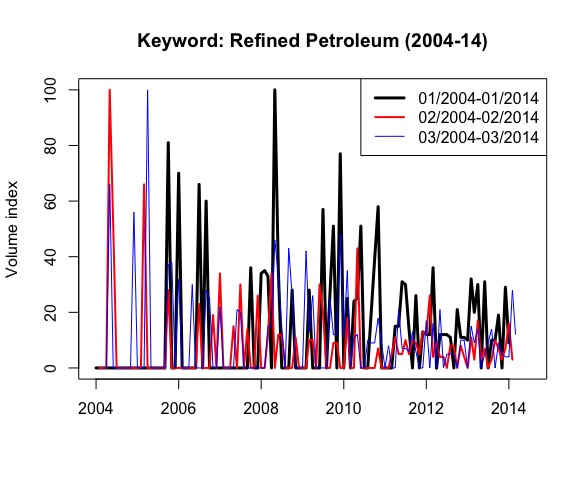}
        \caption{Vintages of ``Refined Petroleum'' topic in the US.}
    \label{fig:rt_refpet_0414}
\end{figure}

It is very clear from Figure \ref{fig:rt_refpet_0414} that each series differ a lot from each other. As a matter of fact, the highest correlation between these series is 0.27 (between black and blue lines) and the lowest is 0.01! It is important to remember that the values of the volume index displayed by \texttt{Google Trends} is normalized and therefore could indeed differ from one vintage to another if the period of search was different. However, it is clear from the plot above that no new maximum search volume index is achieved neither in February, 2014 nor in March, 2014, i.e., the maximum value (100) of the three plots should be the same, when it is not.

Another important thing to mention is that we are showing a very extreme example for two main reasons: 1) we went back very further in time, when fewer people used Google, which means their database should be less consistent; 2) we searched for a very-low volume term, contributing to the low consistency of the data. However, if one intends to perform real-time analysis with \texttt{Google Trends}, it is very likely that she will go back that further in time. Besides that, depending on what she is trying to forecast (e.g., volume or value of exports), she may try to use those low-volume terms in \texttt{Google Trends} as covariates in her forecast.

\begin{figure}[H]
    \centering
    \includegraphics[scale=0.5]{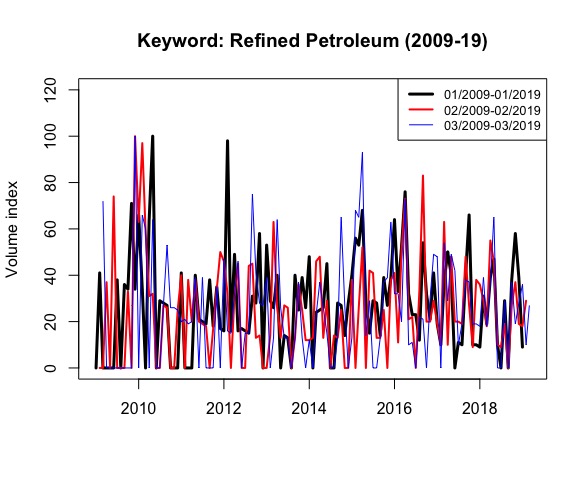}
        \caption{More recent vintages of 'Refined Petroleum' topic in the US.}
    \label{fig:rt_refpet_0919}
\end{figure}

In Figure \ref{fig:rt_refpet_0919} we display the same plot as above, but now with timestamps going from Jan, 2009 - Jan, 2019 to Mar, 2009 - Mar, 2019. We can see that the correlations between each sample becomes a little bit higher: the highest one is one 0.43 and the lowest 0.29.

\begin{figure}[H]
    \centering
    \includegraphics[scale=0.5]{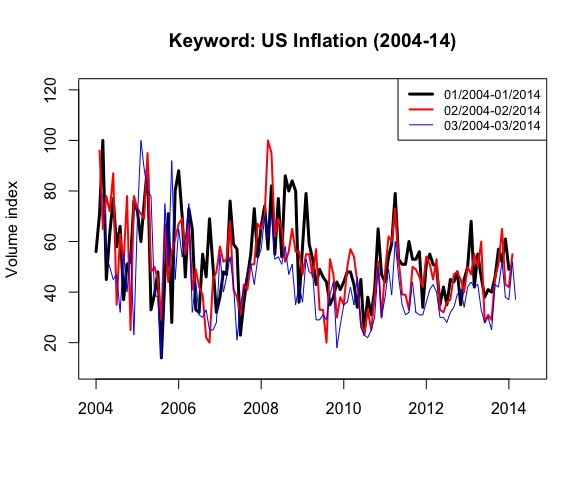}
        \caption{Vintages of 'US Inflation' topic in the US.}
    \label{fig:rt_infl_0414}
\end{figure}

Figure \ref{fig:rt_infl_0414} displays search patterns of ``US inflation'', a far more popular term. We can see that even for the 2004-2014 period, the correlations are already higher (maximum of 0.64 and minimum of 0.44) than the ones from the ``Refined Petroleum'' topic.

It is clear that in recent years correlations can be as high as 0.85. However, this does not mean that taking averages of many samples of these series could not improve their consistency with respect to the actual search pattern of the term in each period of time.

\begin{figure}[H]
    \centering
    \includegraphics[scale=0.5]{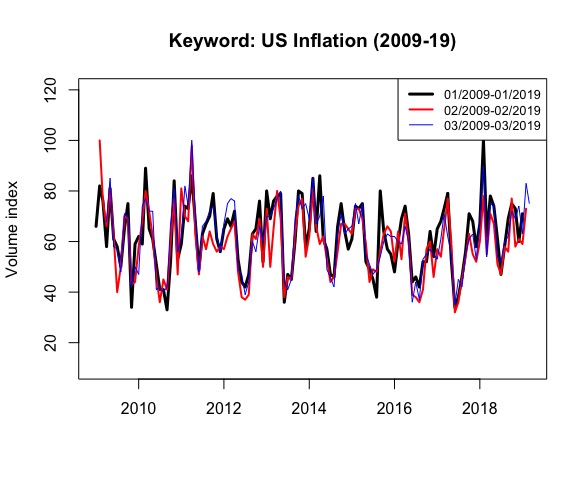}
        \caption{More recent vintages of ``US inflation'' topic in the US.}
    \label{fig:rt_infl_0919}
\end{figure}

\section{Algorithms used in the simulation}\label{sub:appendixB}
\subsection{Setup 1}

\begin{algorithm}
	\caption{}
	\begin{algorithmic}[1]
		\For {$iteration=1,2,\ldots, 1000$}
		\State sample s = a random integer from 1 to 14
		\State sample $\overrightarrow{c}$ = 5 random variables from the 20 available
		\State for k = 1, 2 and 3, construct each $Y^{US}_{k,t,s}$ and each $Y^{BR}_{k,t,s}$ using only variables $\in \overrightarrow{c}$
			\For {$m \in \{1,2 \ldots, 14\} - s$}
				\State Run a LASSO using sample m to predict each $Yk^{BR}_{t,s}$ and $Yk^{US}_{t,s}$
				\State Save which variables were selected by the model to forecast each dependent variable
			\EndFor
		\EndFor
		\State Compute average percentage of variables corrected selected by each LASSO.
	\end{algorithmic}
\end{algorithm}

\subsection{Setup 2}

\begin{algorithm}
	\caption{}
	\begin{algorithmic}[1]
		\For {$iteration=1,2,\ldots, 1000$}
		\State sample $\overrightarrow{s}$ = 7 random integers from 1 to 14 without replacing
		\State for every topic in each sample (20), compute its average across each sample $\in \overrightarrow{s}$
		\State sample $\overrightarrow{c}$ = 5 random variables from the 20 available
		\State for k = 1, 2 and 3, construct each $Yk^{US}_{t,s}$ and each $Yk^{BR}_{t,s}$ using variables $\in \overrightarrow{c}$
		\State Do the same as step 3 but for the other 7 samples $\ni \overrightarrow{s}$
        \State Run a single LASSO using the averaged sample from step 6 to predict each $Y^{BR,US}_{t,s}$
		\State Save which variables were selected by the model to forecast each dependent variable
		\EndFor
		\State Compute average percentage of variables corrected selected by each LASSO.
	\end{algorithmic}
\end{algorithm}
\end{document}